\documentclass[aps,prb,floatfix,epsfig,twocolumn,showpacs,preprintnumbers]{revtex4}
\UseRawInputEncoding
\usepackage{amsfonts}
\usepackage{amssymb}
\usepackage{graphicx,axodraw2}
\usepackage{amsmath}
\usepackage{tablefootnote}
\usepackage{setspace}
\usepackage{caption}
\begin{document}

\title{Full versus quasi-particle self consistency in vertex corrected GW approaches}

\author{Andrey L. Kutepov\footnote{e-mail: akutepov@bnl.gov}}

\affiliation{Condensed Matter Physics and Materials Science Department, Brookhaven National Laboratory, Upton, NY 11973}

\begin{abstract}
Using seven semiconductors/insulators with band gaps covering the range from 1 eV to 10 eV we systematically explore the performance of two different variants of self-consistency associated with famous Hedin's system of equations: the full self-consistency and the so called quasi-particle approximation to it. The pros and cons of these two variants of self-consistency are sufficiently well documented in literature for the simplest GW approximation to the Hedin's equations. Our study, therefore, aims primarily at the level of theory beyond GW approximation, i.e. at the level of theory which includes vertex corrections. Whereas quasi-particle  self-consistency has certain advantages at GW level (well known fact), the situation becomes quite different when vertex corrections are included. In the variant with full self-consistency, vertex corrections (both for polarizability and for self energy) systematically reduce the calculated band gaps making them closer to the experimental values. In the variant with quasi-particle self-consistency, however, an inclusion of the same diagrams has considerably larger effect and calculated band gaps become severely underestimated. Different effect of vertex corrections in two variants of self-consistency can be related to the Z-factor cancellation which plays positive role in quasi-particle self-consistency at GW level of theory but appears to be destructive for the quasi-particle approximation when higher order diagrams are included. Second result of our study is that we were able to reproduce the results obtained with the Questaal code using our FlapwMBPT code when the same variant of self-consistency (quasi-particle) and the same level of vertex corrections (for polarizability only, static approximation for screened interaction, and Tamm-Dancoff approximation for the Bethe-Salpeter equation) are used.
\end{abstract}

\maketitle


\section*{Introduction}
\label{intr}

Reproducibility of results in computational material science is an important issue. In the field of electronic structure calculations, the issue is considered to be resolved at the level of density functional theory (DFT) calculations.\cite{science_351_1415a} General consensus is that modern electronic structure codes, however employ different basis sets (LAPW, LMTO, PAW et ct.), upon convergence demonstrate quite similar electronic structure of materials. When one goes beyond the DFT approximation (for instance if one uses Hedin's GW approach\cite{pr_139_A796}) the number of setup parameters in the calculation increases. Convergence of the results now depends not only on the occupied one-electron states which have to be represented accurately, but also on a number of excited (unoccupied) states which brings considerable difference in the results if excited states are represented differently or if their number (those which included in the calculation) varies. Besides that, the appearance of two-point bosonic functions (such as polarizability P and screened interaction W) requires efficient basis set to represent them. For example, the so called product basis (PB) set\cite{prb_49_16214,prb_99_245149,cpc_184_348} was designed specifically for this purpose. Greater complexity of GW approximation (as compared to DFT) unavoidably brings more differences in implementations which makes the reproducibility of results at the level of GW a more serious issue. Nevertheless, as it was shown in non-self-consistent (G0W0) calculations for 100 molecular systems\cite{jctc_11_5665}, the reproducibility of the molecular electronic structure though not perfect is still acceptable in many cases. Namely, by comparing G0W0 approach implemented in three different codes FHI-aims\cite{njp_14_053020,cpc_180_2175}, BerkeleyGW\cite{cpc_183_1269}, and TURBOMOLE\cite{jctc_9_232,turbomole}, authors of Ref. [\onlinecite{jctc_11_5665}] conclude that at convergence, the HOMO (Highest Occupied Molecular Orbital) and LUMO (Lowest Unoccupied Molecular Orbital) levels agree on the order of 200 meV. In the process of their work, authors of Ref. [\onlinecite{jctc_11_5665}] also have identified two crucial aspects that control the accuracy of the G0W0 quasi-particle energies: the size of the basis set and the treatment of the frequency dependence.

Vertex corrected (diagrammatically) GW calculations for realistic materials now only begin to appear\cite{prl_112_096401,arx_2106_05759,arx_2106_09137,arx_2106_06564,prb_100_045130,prm_5_083805,prb_94_155101,prb_95_195120,prb_96_035108,prb_104_085109,prm_5_083805,arx_2106_03800}.
Their increased complexity (even comparative to G0W0) as well as their relatively recent introduction to the field make the reproducibility of results an open and important issue. Additional (as compared to GW) setup parameters for the methods which diagrammatically go beyond GW approximation are the following: i) specific set of diagrams beyond GW; ii) details of implementation of these high order diagrams. Thorough investigation of the effects of using different sets of diagrams as well as the details of implementation is far beyond the scope of a single study. The objective of this work is more specific and it was motivated by recent vertex corrected GW calculations of the band gaps in semiconductors performed with the FlapwMBPT code\cite{prb_95_195120} and with the Questaal code (B.~Cunningham et al., [\onlinecite{arx_2106_05759}]).

In both studies, [\onlinecite{prb_95_195120}] and [\onlinecite{arx_2106_05759}], it is concluded that vertex corrections provide the biggest numerical improvement of the results obtained with fully self-consistent GW (scGW, Ref. [\onlinecite{prb_95_195120}]) or with quasi-particle self-consistent GW (QSGW, Ref. [\onlinecite{arx_2106_05759}]). Authors of work [\onlinecite{arx_2106_05759}] also include the effect of electron-phonon interaction (for polar semiconductors) but this effect quantitatively is smaller than vertex corrections in most cases. What is important for the present study is the fact that inclusion of electron-phonon interaction always reduces the calculated band gaps as it is evident not only from Ref. [\onlinecite{arx_2106_05759}] but also from earlier works [\onlinecite{prb_89_214304,prl_112_215501,rmp_77_1173,prb_93_100301}]. The list of materials studied in Ref. [\onlinecite{arx_2106_05759}] is slightly different (and also longer) than the corresponding list in Ref. [\onlinecite{prb_95_195120}]. But remarkable tendency of improvements of scGW (or QSGW)  results is unmistakable in both studies. Only one noticeable exception is the case of CuCl in Ref. [\onlinecite{arx_2106_05759}] where vertex correction worsens the QSGW result (band gap becomes seriously underestimated). What is surprising (and which is one of the motivations for the present study) is the fact that similar (and good) results were obtained with quite different variants of vertex corrections applied in two studies.

Whereas both works formally are based on exact Hedin's equations\cite{pr_139_A796}, the details of the applied approximations differ a lot. In the study [\onlinecite{prb_95_195120}], vertex corrections are used to improve both polarizability and self energy $\Sigma$. Authors of Ref. [\onlinecite{arx_2106_05759}] use vertex correction only for polarizability. Further, all vertex corrections in [\onlinecite{prb_95_195120}] (for P and for $\Sigma$) use fully frequency dependent screened interaction $W(\nu)$, whereas vertex correction to P in Ref. [\onlinecite{arx_2106_05759}] is evaluated with frequency independent (taken at zero frequency) $W(\nu=0)$. Also, the Tamm-Dancoff (TD) approximation\cite{nanolett_9_2820} was used in Ref. [\onlinecite{arx_2106_05759}]. There are also some differences in the basis set (see below). But the most important and dramatic (as it will be shown below) difference consists in using full or quasi-particle Green's function in the evaluation of diagrams. In Ref. [\onlinecite{prb_95_195120}] all vertex corrected calculations are performed with full self-consistency applied to Green's function $G$. At the same time, authors of work [\onlinecite{arx_2106_05759}] use additional (the so called quasi-particle, QP) approximation which is not intended in Hedin's equations. Validity of quasi-particle approximation is well justified at the level of GW method (without vertex corrections). It is known, that QSGW approach usually is more accurate than fully self-consistent scGW method.\cite{prb_98_155143,prb_95_195120} The success of QP approximation at the GW level of theory is based on the so called Z-factor cancellation which was clearly explained in the pioneer work on QSGW in Ref. [\onlinecite{prb_76_165106}]. Briefly, the essence of the trick is that diagrams should be evaluated either with full Green's function (no QP approximation) and including vertex part or with QP Green's function but excluding the vertex part. Therefore, if one excludes vertex part (GW level of theory) than it is of advantage to use QP approximation. Alternatively, if one intends to apply vertex corrections, QP approximation for G should not be used and one has to use full G in the evaluation of diagrams instead. Thus, from the point of view of Z-factor cancellation, simultaneous use of vertex corrections and QP approximation for G should be questioned for consistency. Some results on the inadequacy of quasi-particle self-consistency were published a few yeas ago using Hubbard-dimer model as an example\cite{jcm_27_315603}.

The wish to understand deeper the reason why two seemingly different approximations used in [\onlinecite{prb_95_195120}] and [\onlinecite{arx_2106_05759}] lead to similar results was the main motivation for the present work. In order to accomplish the goal we first answer the question whether one can reproduce the results obtained in Ref. [\onlinecite{arx_2106_05759}] (with the Questaal code) using the same approximations as in Ref. [\onlinecite{arx_2106_05759}] but running different code (FlapwMBPT). Secondly, we extend both studies, [\onlinecite{prb_95_195120}] and [\onlinecite{arx_2106_05759}], by performing the calculations which are intended to trace step by step the differences in the implementations of vertex corrections between Refs. [\onlinecite{prb_95_195120}] and [\onlinecite{arx_2106_05759}]. Namely, in the first set of calculations (for each material considered) we start with scGW, then we add vertex correction to polarizability but with $W(\nu=0)$ in the corresponding diagrams, then we apply vertex correction again to P only but with full $W(\nu)$, and finally we add vertex correction to self energy to mimic the full approximation used in Ref. [\onlinecite{prb_95_195120}].

Second set of calculations consists of exactly the same steps but all calculations in the second set are supplemented with QP approximation for Green's function. In this case the variant with vertex correction to P only and with static $W(\nu=0)$ mimics the level of approximation accepted in Ref. [\onlinecite{arx_2106_05759}]. As authors of work [\onlinecite{arx_2106_05759}] also used the Tamm-Dancoff approximation\cite{nanolett_9_2820} in their evaluation of vertex correction to polarizability, the calculations with QP self-consistency (and with static $W(\nu=0)$ approximation) in present work were performed in both ways: with Tamm-Dancoff approximation and without it. In this respect, all calculations with full self-consistency were performed without using the Tamm-Dancoff approximation. Additional steps, i.e. vertex correction to P with full $W(\nu)$ and, finally, with inclusion of vertex correction to $\Sigma$ represent the steps which authors of work [\onlinecite{arx_2106_05759}] mention as possible ways to improve their results but they do not perform these steps. However, as it will be shown below, inclusion of these steps in the calculations with QP approximation for G, in fact, worsens the QP-based results considerably and, therefore, cannot be considered as a valid improvement for calculations with QP self-consistency.

As it is shown below, the calculations performed at similar level of approximations result in very similar results when one uses Questaal code or FlapwMBPT code. At the same time, it is also shown below that a few omissions (or rather 'constraints') accepted in Ref. [\onlinecite{arx_2106_05759}], namely:\begin{itemize}
\item Tamm-Dancoff approximation;
\item insufficient basis set (number of unoccupied states included in vertex corrections);
\item vertex corrections are applied to polarizability only, but not to self energy;
\item static approximation for screened interaction with its value taken at zero frequency;
\end{itemize}
all result in reduction of the calculated band gaps. When the above 'constraints' are removed (with QP self-consistency) the obtained band gaps demonstrate dramatic level of underestimation of the corresponding experimental values. When all 'constraints' are lifted, the calculations even become unstable for small gap semiconductors (when QP self-consistency is used).

The paper begins with a discussion of the approximations used in this study (Section 'Methods'). The discussion of convergence issues and of setup parameters is provided next. The principal results obtained in this work are presented in Section 'Results'. The conclusions are given afterwards.

\section*{Methods}\label{meth}

All calculations in this study formally are based on the Hedin equations.\cite{pr_139_A796} For convenience, we remind the reader about how Hedin's equations could be solved self-consistently in practice.

Suppose one has a certain initial approach for Green's function $G$ and screened interaction $W$. Then one calculates the following quantities:

three-point vertex function from the Bethe-Salpeter equation

\begin{align}\label{Vert_0}
\Gamma^{\alpha}(123)&=\delta(12)\delta(13)\nonumber\\&+\sum_{\beta}\frac{\delta \Sigma^{\alpha}(12)}{\delta
G^{\beta}(45)}G^{\beta}(46)\Gamma^{\beta}(673) G^{\beta}(75),
\end{align}
where $\alpha$ and $\beta$ are spin indexes, and the digits in the brackets represent space-Matsubara's time arguments,

polarizability

\begin{equation}\label{def_pol1}
P(12)=\sum_{\alpha}G^{\alpha}(13)\Gamma^{\alpha}(342)G^{\alpha}(41),
\end{equation}

screened interaction

\begin{align}\label{def_W2}
W(12)=V(12) +V(13)P(34)W(42),
\end{align}

and the self energy

\begin{equation}\label{def_M8}
\Sigma^{\alpha}(12)= - G^{\alpha}(14)\Gamma^{\alpha}(425)W(51).
\end{equation}

In the equation (\ref{def_W2}) V stands for the bare Coulomb interaction.
New approximation for the Green function is obtained from Dyson's equation

\begin{align}\label{D4}
G^{\alpha}(12)=G_{0}^{\alpha}(12) +G_{0}^{\alpha}(13)\Sigma^{\alpha}(34)G^{\alpha}(42),
\end{align}
where $G_{0}$ is the Green function in Hartree approximation. Eqn. (\ref{Vert_0}-\ref{D4}) comprise one iteration. If convergence is not yet reached one can go back to the equation (\ref{Vert_0}) to
start the next iteration with renewed $G$ and $W$.

The system of Hedin's equations formally is exact, but one has to introduce certain approximations when solving (\ref{Vert_0}) for the vertex function $\Gamma^{\alpha}(123)$ in order to make the solving of the system manageable in practice. Approximations which we use in this study are dictated by the goals of the work. In order to justify their choice let us summarize the goals again:\begin{itemize}
\item reproduce the results of Ref. [\onlinecite{arx_2106_05759}];
\item check the validity of speculations made in Ref. [\onlinecite{arx_2106_05759}] about the possibility to improve QSGW-based results by taking into account frequency dependence of W when solving the BSE and adding vertex correction to self energy;
\item compare full self-consistency with QP self-consistency.
\end{itemize}

Guided by the goals, we conducted two sets of calculations (for each material studied) which were already sketched in the Introduction section. Starting point for the first set is scGW approximation (fully self-consistent GW), whereas second set of calculations has QSGW approach as a starting point (as in Ref. [\onlinecite{arx_2106_05759}]). In scGW method, vertex function (\ref{Vert_0}) is approximated by its trivial part $\Gamma^{\alpha}(123)=\delta(12)\delta(13)$. In QSGW, we use additional approximation related to the self energy in Eq. (\ref{D4}), namely we linearize frequency dependence of self energy around zero frequency (see details in Refs. [\onlinecite{prb_85_155129,cpc_219_407}]). In this respect, our construction of QSGW differs from Ref. [\onlinecite{arx_2106_05759}] where the approximation for self energy in Eq. (\ref{D4}) consists in taking its hermitian part. As it will be shown below, however, numerical results obtained with above two variants of QSGW are pretty much similar for majority of materials. Concerning the vertex part, all vertex corrected calculations included the solution of Bethe-Salpeter equation (BSE) for polarizability with screened interaction W as a kernel of BSE ($\frac{\delta \Sigma^{\alpha}(12)}{\delta
G^{\beta}(34)}\approx \delta_{\alpha\beta}\delta(13)\delta(24)W(12)$). Corresponding diagrammatic representation for the vertex correction to P is shown in Fig. \ref{bse}. At this level, we conducted a few variants of calculations in order to explore approximations made by authors of Ref. [\onlinecite{arx_2106_05759}], namely the Tamm-Dancoff approximation and the use of frequency independent $W(\nu=0)$. Finally, our the most sophisticated calculations included vertex correction to self energy of second order (second term in Fig. \ref{diag_S}). This approximation for self energy corresponds to expansion of vertex function in Eq. (\ref{Vert_0}) up to the first order in W. As one can easily notice, the above described vertex corrected variants assume different approaches for vertex function when it is used in the expression for polarizability (\ref{def_pol1}) and in the expression for self energy (\ref{def_M8}) and, as a result, they are not conserving in Baym-Kadanoff definition\cite{pr_124_287} (i.e. corresponding P and $\Sigma$ cannot be obtained from the same functional). In order to check the effect of using a scheme which is conserving, we also included vertex corrected calculations (with full self-consistency only, not QP) where vertex correction to polarizability consists of the first only term shown in Fig. \ref{bse} and vertex correction to self energy consists of second term in Fig. \ref{diag_S}. This approach, sc(GW+G3W2), as well as scGW, can also be defined using $\Psi$-functional formalism of Almbladh et al.\cite{ijmpb_13_535} Corresponding $\Psi$-functional which includes vertex corrections is shown in Fig. \ref{diag_Psi}. In Fig. \ref{diag_Psi}, the first diagram corresponds to GW approximation, whereas the sum of the first and the second diagram represents sc(GW+G3W2) approximation. Diagrammatic representations for irreducible polarizability (Fig. \ref{diag_P}) and for self energy (Fig. \ref{diag_S}) in sc(GW+G3W2) follow from the chosen approximation for $\Psi$-functional.

For convenience, we list here all variants of approximations used in this study together with the purpose and with the corresponding abbreviations. First set of calculations includes:\begin{itemize}
\item scGW, which is used primarily to generate initial approximation to start vertex corrected calculations and, by doing this, to reduce number of iterations with vertex corrections. We also compare the scGW results with the ones obtained with QSGW;
\item sc(BSE0:P$@$GW), where the part after the symbol $@$ stands for diagrammatic representation of self energy, whereas the part before the symbol $@$ says that polarizability is evaluated from BSE with static screened interaction taken at zero frequency ($W(\nu=0)$) as the kernel of BSE. The goal of this variant is to assess the quality of static approximation for W in BSE, as well as to compare this variant with the same variant but based on QP self-consistency;
\item sc(BSE:P$@$GW), where one uses full frequency dependent $W(\nu)$ in BSE. The goal of this variant is to assess the quality of static approximation for W in BSE (by comparing results with sc(BSE0:P$@$GW)) and also to compare this variant with the same variant but based on QP self-consistency;
\item sc(BSE:P$@$GW+G3W2), where diagrammatic representation of self energy includes second oreder (in W) diagram. All diagrams in this variant (for P and for $\Sigma$) use full frequency dependent $W(\nu)$. The goal of this variant is to assess the effect of inclusion of vertex correction to self energy and also to compare this variant with the same variant but based on QP self-consistency;
\item sc(GW+G3W2), which is conserving in Baym-Kadanoff definition. Only diagrammatic definition of self energy (GW+G3W2) is needed to be specified in this case. Diagrammatic representation for P, G2+G4W1, follows if ones uses the same first order vertex function (as for self energy) in Eq. (\ref{def_pol1}). All diagrams in this variant (for P and for $\Sigma$) also use full frequency dependent $W(\nu)$. The goal of this variant is to assess the effect of applying the conserving approximation.
\end{itemize}

Second set of calculations includes:\begin{itemize}
\item QSGW, which is used to compare the results with the ones obtained in Ref. [\onlinecite{arx_2106_05759}] at the same level of theory;
\item QPsc(BSE0TD:P$@$GW), where the part in brackets before the symbol $@$ says that polarizability is evaluated from BSE with static screened interaction taken at zero frequency ($W(\nu=0)$) as the kernel of BSE. Plus, Tamm-Dancoff approximation is assumed. The basis size for BSE is taken exactly as in Ref. [\onlinecite{arx_2106_05759}]. Initial symbols 'QP' stand for quasi-particle self-consistency. The goal of this variant is to compare the results with the ones obtained in Ref. [\onlinecite{arx_2106_05759}] at the same level of theory and to assess the effect of Tamm-Dancoff approximation;
\item QPsc(BSE0:P$@$GW), which is the same as sc(BSE0:P$@$GW) but with QP self-consistency instead of full. The goal of this variant is to assess the effect of Tamm-Dancoff approximation and of the differences in size of the basis set for BSE;
\item QPsc(BSE:P$@$GW), which is the same as sc(BSE:P$@$GW) but with QP self-consistency instead of full. The goal of this variant is to assess the quality of static approximation for W in BSE (by comparing results with QPsc(BSE0:P$@$GW) in the case of QP self-consistency;
\item QPsc(BSE:P$@$GW+G3W2), which is the same as sc(BSE:P$@$GW+G3W2) but with QP self-consistency instead of full. The goal of this variant is to assess the effect of inclusion of vertex correction to self energy when one uses QP self-consistency;
\end{itemize}

\begin{figure}[t]
\begin{center}\begin{axopicture}(200,56)(0,0)
\SetPFont{Arial-bold}{28}
\SetWidth{0.8}
\Text(-26,10)[l]{$\Delta P$  = -}
\Photon(13,10)(17,10){2}{1.5}
\GCirc(37,10){20}{1}
\Photon(37,-10)(37,30){2}{5.5}
\Photon(57,10)(61,10){2}{1.5}
\Text(66,10)[l]{$+$}
\Photon(78,10)(82,10){2}{1.5}
\GCirc(102,10){20}{1}
\Photon(95,-8)(95,28){2}{4.5}
\Photon(109,-8)(109,28){2}{4.5}
\Photon(122,10)(126,10){2}{1.5}
\Text(131,10)[l]{$-$}
\Photon(144,10)(148,10){2}{1.5}
\GCirc(168,10){20}{1}
\Photon(158,-6.5)(158,26.5){2}{4.5}
\Photon(168,-10)(168,30){2}{4.5}
\Photon(178,-6.5)(178,26.5){2}{4.5}
\Photon(188,10)(192,10){2}{1.5}
\Text(195,10)[l]{$+ ...$}
\end{axopicture}
\end{center}
\caption{Ladder sequence of diagrams for the vertex correction to polarizability.}
\label{bse}
\end{figure}

\begin{figure}[t]
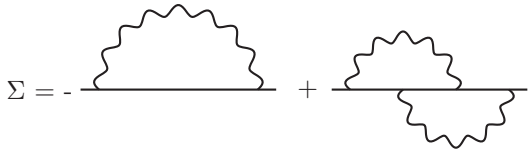

\begin{center}\begin{axopicture}(200,56)(0,0)
\SetPFont{Arial-bold}{28}
\SetWidth{0.8}
\Text(10,10)[l]{$\Sigma$  = -}
\Line(38,10)(112,10)
\PhotonArc(75,10)(30,0,180){2}{8.5}
\Text(120,10)[l]{$+$}
\Line(133,10)(207,10)
\PhotonArc(160,10)(20,0,180){2}{6.5}
\PhotonArc(180,10)(20,180,360){2}{6.5}
\end{axopicture}
\end{center}
\caption{Diagrammatic representation of self energy up to the second order in screened interaction W.}
\label{diag_S}
\end{figure}

\begin{figure}[t]
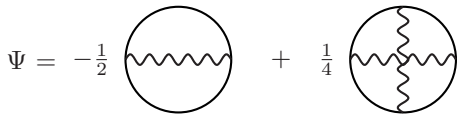

\begin{center}\begin{axopicture}(200,56)(0,0)
\SetPFont{Arial-bold}{28}
\SetWidth{0.8}
\Text(10,10)[l]{$\Psi$ =}
\Text(35,10)[l]{$-\frac{1}{2}$}
\GCirc(75,10){20}{1}
\Photon(55,10)(95,10){2}{5.5}
\Text(110,10)[l]{+}
\Text(128,10)[l]{$\frac{1}{4}$}
\GCirc(160,10){20}{1}
\Photon(160,-10)(160,30){2}{5.5}
\Photon(140,10)(180,10){2}{5.5}
\end{axopicture}
\end{center}
\caption{Diagrammatic representation of $\Psi$-functional which includes the simplest non-trivial vertex. First diagram on the right hand side stands for scGW approximation, whereas total expression corresponds to sc(GW+G3W2) approximation.}
\label{diag_Psi}
\end{figure}

\begin{figure}[t]
\begin{center}\begin{axopicture}(200,56)(0,0)
\SetPFont{Arial-bold}{28}
\SetWidth{0.8}
\Text(10,10)[l]{$P$  =}
\Photon(48,10)(55,10){2}{2.5}
\GCirc(75,10){20}{1}
\Photon(95,10)(102,10){2}{2.5}
\Text(120,10)[l]{$-$}
\Photon(143,10)(150,10){2}{2.5}
\GCirc(170,10){20}{1}
\Photon(190,10)(197,10){2}{2.5}
\Photon(170,-10)(170,30){2}{5.5}
\end{axopicture}
\end{center}
\caption{Diagrammatic representation of irreducible polarizability in the simplest conserving vertex corrected scheme sc(GW+G3W2).}
\label{diag_P}
\end{figure}

\begin{figure}[t]
\begin{center}       
\includegraphics[width=8.5 cm]{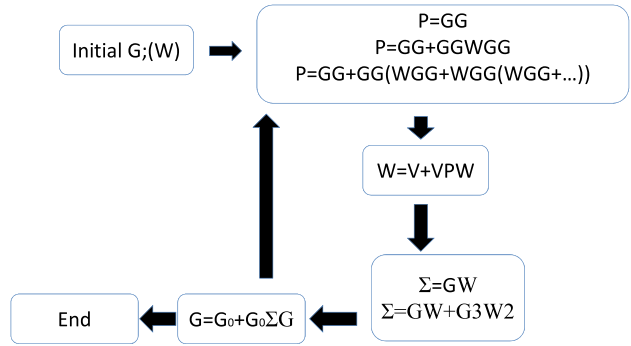}
\caption{Flowchart of scGW, sc(GW+G3W2), and sc(BSE:P$@$GW+G3W2) calculations. All equations are presented using symbolic notations. In the expressions for polarizability, first equation corresponds to scGW, second equation is used in sc(GW+G3W2), and the third one in sc(BSE:P$@$GW+G3W2). In the expressions for self energy, first equation corresponds to scGW, and the second one to both sc(GW+G3W2) and sc(BSE:P$@$GW+G3W2). $G_{0}$ stands for Green's function in Hartree approximation. Any calculation begins with self-consistent DFT iterations where the basis set is formed and the initial approach for G is generated. Iterations of scGW method use this initial Green's function as an input in order to start. During scGW iterations, G is updated and screened interaction W is generated. Both G and W serve as an input to start iterations of sc(GW+G3W2) or sc(BSE:P$@$GW+G3W2) approaches. sc(BSE:P$@$GW+G3W2), being computationally most demanding, can be run after a few iterations of sc(GW+G3W2), which can save computer time.}
\label{flow}
\end{center}
\end{figure}

All calculations in this work were performed using code FlapwMBPT.\cite{flapwmbpt_1} Technical details of the GW part were described in Refs. [\onlinecite{prb_85_155129,cpc_219_407}]. Detailed account of the implementation of vertex corrected schemes can be found in Refs. [\onlinecite{prb_94_155101,prb_95_195120,prb_96_035108,prb_104_085109}]. Figure \ref{flow} presents the flowchart of the calculations (for three selected approximations) which gives a general idea of how the calculations are organized. The flowchart in Fig. \ref{flow} corresponds to full self-consistency. In the case of quasi-particle self-consistency, the formal change consists only in the fact that instead of Dyson's equation ($G=G_{0}+G_{0}\Sigma G$), a special construction for $G$ is used as it is described in Refs. [\onlinecite{prb_85_155129,cpc_219_407}]. The diagrammatic (GW and the diagramms beyond GW) parts of the FlapwMBPT code take full advantage of the fact that certain diagrams can more efficiently be evaluated in reciprocal (and frequency) space whereas other diagrams are easier to evaluate in real (and time) space. As a result, GW part of the code scales as $N_{k}N_{\omega}N^{3}_{b}$ where $N_{k}$ is the number of \textbf{k}-points in the Brillouin zone, $N_{\omega}$ is the number of Matsubara frequencies, and $N_{b}$ stands for the size of the basis set. The vertex part of the code scales as $N^{2}_{k}N^{2}_{\omega}N^{4}_{b}$. For comparison, if one uses naive (all in reciprocal space and frequency) implementation then GW part scales as $N^{2}_{k}N^{2}_{\omega}N^{4}_{b}$ (i.e. exactly as the vertex part when the implementation is efficient), and the vertex part scales as $N^{3}_{k}N^{3}_{\omega}N^{5}_{b}$. Besides of efficiency of the implementation, we have to mention two more factors which make the use of the diagrams beyond GW feasible. First is the fact that the higher order diagrams converge much faster than the GW diagram with respect to the basis set size and to the number of \textbf{k}-points.\cite{prb_94_155101,prb_95_195120} Second is that the higher order diagrams are very well suited for massive parallelization.

\section*{Calculation setups and convergence checks}\label{setup}

Let us now specify the setup parameters used in the calculations. First of all, our selection of materials for this study was dictated by the following constraints: i) band gaps of the selected compounds should cover (approximately uniformly) a broad range of energies (1 -- 10 eV); ii) selected materials should be taken from the list studied in Ref. [\onlinecite{arx_2106_05759}] (only $AlP$ does not fit in this constraint); iii) selected materials should be sufficiently simple as our vertex corrected calculations which use full frequency dependence of W are rather time consuming. In order to make presentation more compact, the list of selected compounds, their principal structural parameters and the size of basis sets have been collected in Table \ref{list_s}. All calculations have been performed for the electronic temperature $600K$ (Matsubara's formalism is used throughout the work). Commonly used setup parameter $RKmax$ for LAPW-based calculations was set to 8.0 in all calculations of this work. The sampling of the Brillouin zone for GW part (i.e. excluding vertex correction diagrams) was $12\times 12\times 12$ in all cases. Evaluation of the diagrams associated with vertex part (i.e. all diagrams in Fig. \ref{bse} and the second diagram in Fig. \ref{diag_S}) was performed with sampling $3\times 3\times 3$. As Table \ref{list_s} shows, the number of band states used in the evaluation of vertex part also was considerably smaller than the number of band states included in the evaluation of GW part of the diagrams. The fact, that the diagrams representing the vertex part require smaller basis set and coarser sampling of the Brillouin zone was discussed before in Refs. [\onlinecite{prb_94_155101,prb_95_195120}], so that the choice of these two setup parameters for the present study is justified (see for instance Table I in Ref. [\onlinecite{prb_95_195120}]). At the same time, one can notice that the basis set for the BSE part used in Ref. [\onlinecite{arx_2106_05759}] is smaller than ours by almost a factor of two (our basis sets for the vertex part shown in Table \ref{list_s} are the sums of valence and conduction bands included). This fact was the reason that we perform our QPsc(BSE0TD$@$GW) calculations with the basis set (for the vertex part) exactly corresponding to the basis set used in Ref. [\onlinecite{arx_2106_05759}]. Therefore, the difference between the band gaps obtained with QPsc(BSE0$@$GW) and QPsc(BSE0TD$@$GW) is, in fact, a total effect of the TD approximation and of the basis set mismatch. In most cases, however, the effect of TD approximation was prevailing. 

\begin{table}[t]
\caption{Setup parameters of the solids studied in this work. Lattice parameters are in Angstroms. $N_{bnd}^{GW}$ is the number of band states used as a basis set for evaluation of GW part. $N_{bnd}^{VRT}$ represents the corresponding number for the vertex part.} \label{list_s}
\small
\begin{center}
\begin{tabular}{@{}c c c c c} &Space&Lattice&&\\
Solid &group&parameter&$N_{bnd}^{GW}$&$N_{bnd}^{VRT}$\\
\hline\hline
Si&227 &5.43&160&20\\
AlP&216 &5.451&185&14\\
CuCl&216 &5.64&260&24\\
C&227 &3.57&160&14\\
MgO&225 &4.217&110&16\\
NaCl&225 &5.62&150&16\\
LiCl&225 &5.13&120&16\\
\end{tabular}
\end{center}
\end{table}

It is well known that LAPW basis set has to be supplemented with sufficient number of high energy local orbitals (HELO) in order to ensure the convergence of calculated band gaps in GW-based approximations (see for instance Refs. [\onlinecite{prb_74_045104,prb_83_081101,prb_84_039906,prb_93_115203,prb_94_035118}]). Therefore, for all studied materials, we extended standard LAPW basis set by including 3--4 (per atom) HELO's of s-type, 2--3 HELO's of p- and d-type, 1--2 HELO's of f-type, and also 1 HELO of g-type. In this respect, our additional basis set (HELO's) was also larger than additional local orbital basis set used in Ref. [\onlinecite{arx_2106_05759}] (see Table I there) which can be another reason for small differences in results at QSGW level (besides of the different way to introduce the quasi-particle approximation).

In most of our vertex corrected calculations (excluding QPsc(BSE0TD$@$GW)), BSE is solved iteratively\cite{prb_94_155101} which especially is needed when one uses frequency dependent $W(\nu)$. Referring to Fig. \ref{bse}, one iteration in this case means adding one more term in the infinite sequence of ladder diagrams. In practice, the infinite sequence has to be truncated. In this study we used six terms as a cutoff parameter for the iterative solution of BSE. As we did show before (see Fig. 7 in Ref. [\onlinecite{prb_95_195120}]), this choice of the cutoff means that the contribution of the rest of ladder diagrams (i.e. those which are not included) could be only 1/50--1/100 of the first term contribution. In fact, our checks with eight terms have demonstrated that the addition of two more ladder diagrams changes the calculated band gaps by less than 0.005 eV.

Our QPsc(BSE0TD$@$GW) approach which serves to reproduce the results obtained in Ref. [\onlinecite{arx_2106_05759}] was implemented similar to as described in Ref. [\onlinecite{arx_2106_05759}]. Namely, we solved BSE directly (not iteratively as in all other our approaches) using \textbf{k}-dependent product basis for electron-hole pairs. Details of this basis set specific for LAPW implementation can be found in Ref. [\onlinecite{elstr_1_037001}]. Tamm-Dancoff approximation, therefore, was implemented by neglecting the anti-resonant part and keeping only the resonant part of the transition space\cite{elstr_1_037001}.

As it will be shown later in this work, the approximation of full frequency dependent $W(\nu)$ by static function $W(\nu=0)$ gives qualitatively (but not quantitatively) correct results for vertex corrections to polarizability. However, as it was discussed before\cite{prb_94_155101}, similar replacement of $W(\nu)$ by $W(\nu=0)$ in the vertex correction for self energy gives even qualitatively incorrect results. Namely, vertex correction to the band gaps evaluated with static $W(\nu=0)$ is positive (band gaps increase) whereas correct evaluation of the corresponding diagrams (with full $W(\nu)$) always reduces the gaps. This observation also collaborates with the increase in the calculated band gaps obtained by A.~Gr\"{u}neis et al. in Ref. [\onlinecite{prl_112_096401}] where static $W(\nu=0)$ was used. By this reason, self energy vertex corrections in this work were always evaluated with full frequency dependent $W(\nu)$.

\section*{Results}
\label{res}

\begin{figure*}[t] 
\begin{center}       
    \includegraphics[width=12.5 cm]{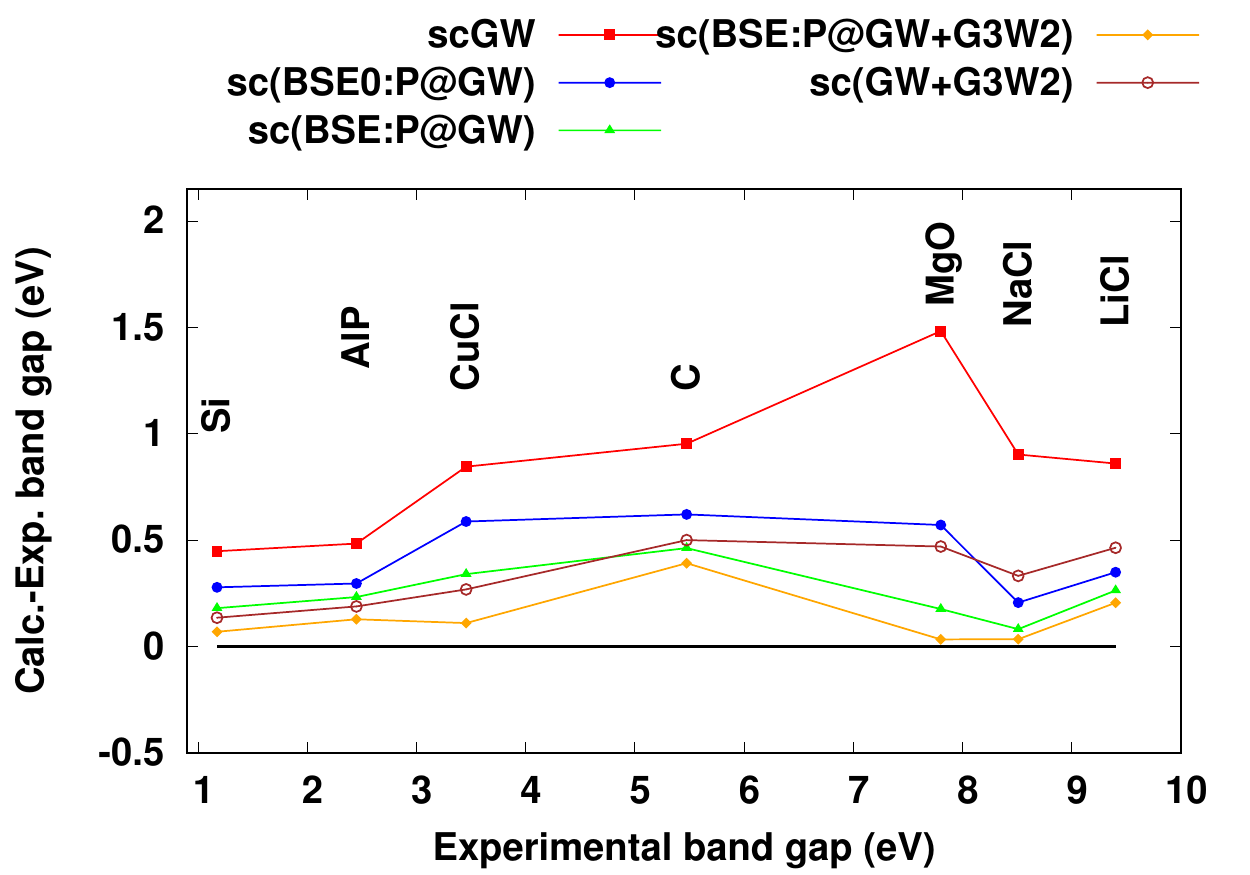}
    \caption{Band gaps as obtained with full self-consistency. Positioning of the data for each compound relative to the X-axis corresponds to the experimental band gap of the specific compound (Refs.[\onlinecite{prb_20_624,prb_76_165106,prb_35_9174,prb_53_16283}]). Y-axis represents the deviations of the calculated band gaps from the experimental ones. Calculated results do not include electron-phonon interaction.}
    \label{gw_gaps}
\end{center}
\end{figure*}

\begin{figure*}[t] 
\begin{center}       
    \includegraphics[width=12.5 cm]{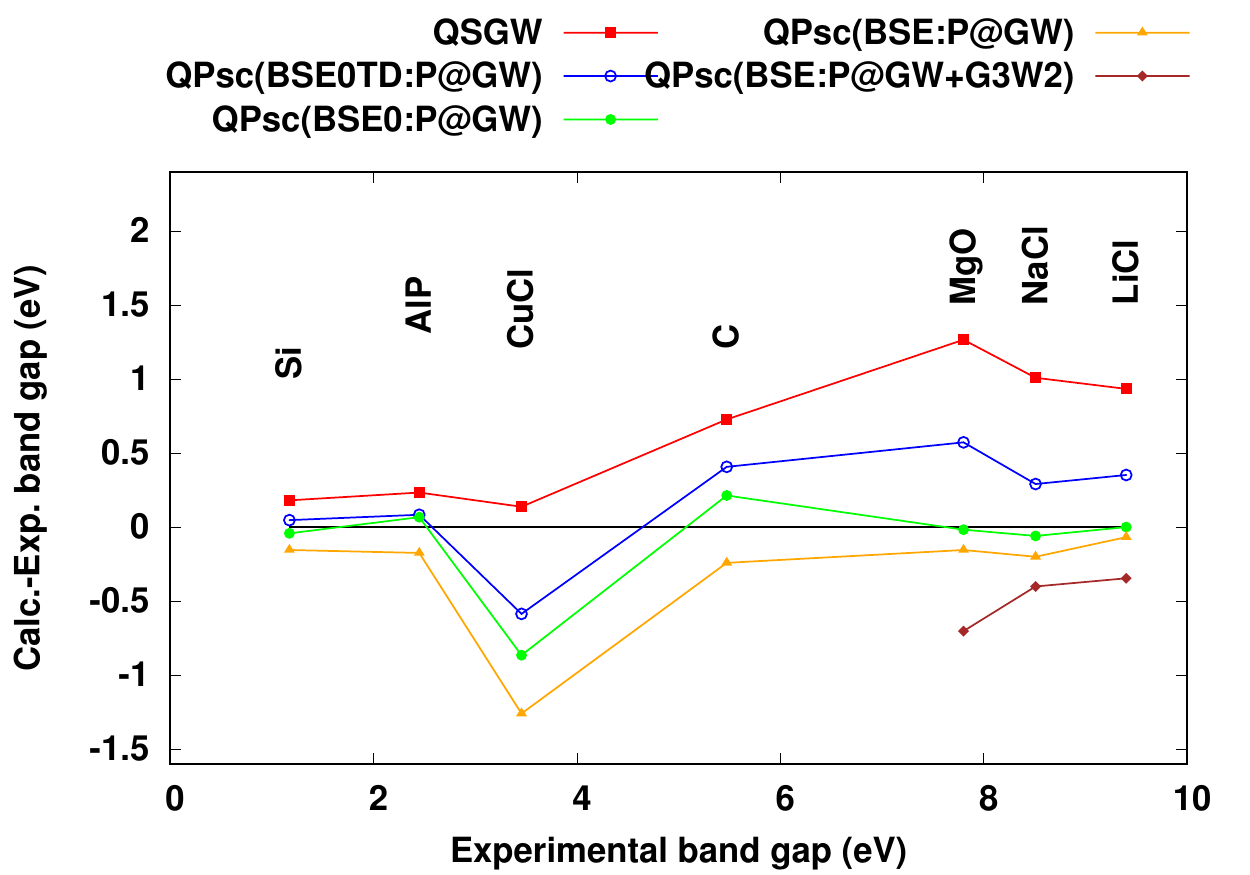}
    \caption{Band gaps as obtained with quasi-particle self-consistency. Positioning of the data for each compound relative to the X-axis corresponds to the experimental band gap of the specific compound (Refs.[\onlinecite{prb_20_624,prb_76_165106,prb_35_9174,prb_53_16283}]). Y-axis represents the deviations of the calculated band gaps from the experimental ones. Calculated results do not include electron-phonon interaction. QPsc(BSE:P$@$GW+G3W2) band gaps are presented only for materials where the self-consistency iterations are stable.}
    \label{qp_gaps}
\end{center}
\end{figure*}

Principal results of this work are presented in Figs. \ref{gw_gaps} and \ref{qp_gaps} where the band gaps obtained with full self-consistency (Fig. \ref{gw_gaps}) and with QP self-consistency (Fig. \ref{qp_gaps}) are shown. Let us start our discussion with the full self-consistency. Firstly, we observe that calculations without vertex corrections (scGW) severely overestimate the calculated band gaps (by 0.5--1.5 eV). Secondly, principal improvement comes from the vertex correction to polarizability as it is evidenced in sc(BSE:P$@$GW) calculations. It is important to point out, however, that in order to get quantitatively correct results one has to use full frequency dependent $W(\nu)$ when solving BSE which is used to obtain vertex correction to polarizability. Using static $W(\nu=0)$ in BSE can only give qualitatively correct correction to P, but quantitatively it underestimates the correction by 20--50\% (compare sc(BSE0:P$@$GW) with sc(BSE:P$@$GW)). Thirdly, vertex correction to self energy always reduces the calculated band gaps (compare sc(BSE:P$@$GW) with sc(BSE:P$@$GW+G3W2)). It needs to be pointed out that we always use full frequency dependent $W(\nu)$ in the evaluation of self energy vertex correction (as we already mentioned before). As it was discussed in Ref. [\onlinecite{prb_94_155101}], the use of static $W(\nu=0)$ results in qualitatively incorrect correction to band gaps: they are increasing instead of decreasing when one uses full $W(\nu)$. The effect of vertex correction to self energy is smaller than the effect of vertex correction to polarizability but it still is important as the total result (band gaps as obtained in sc(BSE:P$@$GW+G3W2)) is very close to the experimental band gaps with remaining discrepancy mostly attributed to the electron-phonon interaction which was not included in present study. For instance, the biggest discrepancy in the band gap of 0.5 eV (case of carbon) can be nicely accounted for by considering the corresponding electron-phonon band reduction ($\sim$ 0.4 eV, [\onlinecite{prb_89_214304,ssc_133_3,prb_90_184302}]). Finally, Fig. \ref{gw_gaps} also includes the band gaps evaluated with conserving (in Baym-Kadanoff\cite{pr_124_287} definition) approach sc(GW+G3W2). This approach has a merit of not only being conserving but also of being more computationally efficient because only one diagram (of first order) in the sequence of Fig. \ref{bse} has to be evaluated for polarizability vertex correction. As one can judge from Fig. \ref{gw_gaps}, sc(GW+G3W2) approach is especially useful for small gaps semiconductors where the corresponding band gaps are close to the calculated with sc(BSE:P$@$GW+G3W2) band gaps. For large gap insulators, however, solving of full BSE for polarizability vertex correction is essential.

Let us discuss now the results obtained with QP self-consistency which are presented in Fig. \ref{qp_gaps}. As one can conclude, our results confirm that at GW level, quasi-particle self-consistency works better than full self-consistency. Especially it is noticeable for small gap (1--3 eV) materials. Next important observation from Fig. \ref{qp_gaps} is that our QPsc(BSE0TD:P$@$GW) results are not only in qualitative but also in quantitative agreement with the corresponding results obtained using the Questaal code (see Fig. 6 in Ref. [\onlinecite{arx_2106_05759}]). Similar to Ref. [\onlinecite{arx_2106_05759}], QPsc(BSE0TD:P$@$GW) approach does a good job in bringing the calculated band gaps in close agreement with experiments (especially if one takes into account the electron-phonon correction as authors of Ref. [\onlinecite{arx_2106_05759}] demonstrate). Similar to the work [\onlinecite{arx_2106_05759}], there is one notable exception, CuCl, where the calculated with QPsc(BSE0TD:P$@$GW) band gap is severely underestimated. Thus, we are arriving to an important conclusion that using the same approximation, QPsc(BSE0TD:P$@$GW), we are able to reproduce the results obtained with the Questaal code in Ref. [\onlinecite{arx_2106_05759}]. It is, therefore, interesting that partial vertex correction (correction to only P, static $W(\nu=0)$, TD approximation) combined with quasi-partical self-consistency mimics the total result (vertex corrections to both P and $\Sigma$, full $W(\nu)$, no Tamm-Dancoff approximation) obtained with full self-consistency. However, in disagreement with the speculations made by authors of Ref. [\onlinecite{arx_2106_05759}] that results can be further improved by using full $W(\nu)$ and also adding vertex correction to self energy, we see from Fig. \ref{qp_gaps} that both speculated 'improvements' result in too big reduction of the calculated band gaps making them seriously underestimated. In fact, when self energy vertex correction is taken into account, calculations become unstable for small gap materials. Thus, as it seems, the consideration about Z-factor cancellation done by authors of work [\onlinecite{prb_76_165106}] works. But, in conjunction with vertex corrections, it works in negative direction essentially explaining the fact that QP self-consistency should not be combined with vertex corrections. One more result, which one can get from Fig. \ref{qp_gaps} is that TD approximation is of rather poor quality (especially for large gap insulators) when one uses it in the context of band gaps evaluation which involves integration over the Brillouin zone. This finding corroborates with the finding made by authors of Ref. [\onlinecite{prb_92_045209}] that TD approximation fails for finite momentum transfers. Summarizing our observations of QP self-consistency, we can state that this variant of self-consistency can only be combined with vertex corrections if one makes additional (which are not assumed in Hedin's equations) approximations such as Tamm-Dancoff approximation, static $W(\nu=0)$, and no self energy vertex correction. As it is now confirmed empirically by two different codes (Questaal and FlapwMBPT) such approximations still allow to improve QSGW band gaps but at the same time allow one to avoid destructive effect of Z-factor cancellation when full vertex corrections are used in connection with QP self-consistency.

\section*{Conclusions}
\label{concl}

The study conducted in the present work resulted in two principal conclusions. The first one, which clearly represents a positive achievement, is that two codes (Questaal and FlapwMBPT) produce similar band gaps in vertex corrected QSGW calculations for a number of materials provided that vertex corrections are evaluated similarly (correction to only polarizability, static $W(\nu=0)$, Tamm-Dancoff approximation for BSE). The second conclusion is that when one adds diagrams (beyond GW) in self-consistent calculations one should use full self-consistency approach. Namely, Green's function has to be properly evaluated from Dyson's equation without referring to the quasi-particle approximation. As it is shown in this study, combining the vertex corrected calculations with QP self-consistency as it is advocated in Refs. [\onlinecite{arx_2106_05759,arx_2106_09137,arx_2106_06564}], can only be successful if vertex corrections are evaluated with a number of restrictive approximations or 'constraints' such as polarizability only correction, static $W(\nu=0)$ in vertex diagrams, and Tamm-Dancoff approximation when solving BSE. From this point of view, quasi-particle self-consistency combined with vertex-corrected GW approach can be considered as an \textit{ad hoc} theory where one imposes specific constraints on the vertex part in order to avoid too large (and destructive for the final result) effect. Nevertheless, the approach still can be useful from practical point of view allowing one to quickly estimate the possible effect of vertex corrections before addressing the problem with full vertex and full self-consistency.

The second conclusion, as it seems, is in a contradiction with our previous advocating the combination of QSGW and dynamical mean field theory (QSGW+DMFT, [\onlinecite{npj_1_16001,cpc_244_277}]. Formally, the addition of DMFT to QSGW can be considered as a vertex correction and, according to the discussion above, cannot be a valid approximation. However, similar to the approach used in Refs. [\onlinecite{arx_2106_05759,arx_2106_09137,arx_2106_06564}], our implementation of QSGW+DMFT also uses 'constraints' for the vertex (DMFT) part: i) only one iteration which includes DMFT (one-shot type of DMFT correction performed on top of QSGW); ii) effective interaction in DMFT part is not evaluated from proper DMFT self-consistency condition, [\onlinecite{prl_90_086402}], but is provided by constrained random-phase approximation (cRPA, [\onlinecite{prb_70_195104}]). Specifically, the second constraint (using the cRPA) can clearly be considered as an \textit{ad hoc} part where the setup parameters of cRPA are adjusted in order to get reasonable effective interaction. Thus, the success of our QSGW+DMFT calculations can also be attributed to the use of 'constraints' in the vertex part.

\section*{Acknowledgments}
\label{acknow}
This work was   supported by the U.S. Department of energy, Office of Science, Basic
Energy Sciences as a part of the Computational Materials Science Program.



\end{document}